\documentclass[aps,prresearch,twocolumn,10pt,superscriptaddress,longbibliography]{revtex4-1}

\usepackage{amsmath}
\usepackage{graphicx}
\usepackage{times}
\usepackage{amssymb}
\usepackage{hyperref}
\usepackage{textcomp}
\usepackage{xcolor}
\usepackage[T1]{fontenc}
\usepackage{ulem}

\begin{document}

\title{Nonquantum Information Gain from Higher-order Correlation Functions}

\author{Peter Gr\"unwald}
\affiliation{Aarhus Universitet, Institut for Fysik og Astronomi, Ny Munkegade 120, 8000 Aarhus C, Denmark.
}
\email{peter.gruenwald@phys.au.dk}

\begin{abstract}
Nonlinear correlation functions are at the heart of quantum theory. The second-order correlation function $g^{(2)}(\tau)$ has been a cornerstone of quantum optics since over half a century and a myriad of quantum and classical applications has been discovered. In contrast, higher-order correlation functions have so far only been used to reveal the nonclassical character of the emitted fields.
In this paper, we study the relation between the $k$th-order correlation function $g^{(k)}(0)$ and the projection of the underlying quantum state of light onto the subspace of Fock states with photon number less than $k$. We show, that when $g^{(k)}(0)$ falls below a critical value, lower bounds for the projection on this subspace can be concluded as well as on the ratio of the subspace with one upto $k-1$ photons and $k$ to infinity. These bounds are at face value only valid for nonclassical quantum states. However, when the quantum state includes a nonzero projection on the vacuum state, the value of $g^{(k)}(0)$ is artificially enhanced, potentially covering these projections. We derive an effective $k$th-order correlation function, which accounts for the effect of vacuum. 
We show that the information gained from the effective correlation function is not limited to nonclassical quantum states and thus constitute a quantum- and classical application of higher-order correlation functions.
\end{abstract}

\maketitle

\section{introduction}\label{sec.intro}
One of the main features of quantum physics, which is readily available to experiments, is the strongly nonlinear character of higher-order expectation values and correlation functions. Already introductory courses on quantum mechanics focus on the intrinsic variance $\langle(\hat X-\langle\hat X\rangle)^2\rangle$ of an observable $\hat X$ being nonzero if the quantum state is not an eigenstate of $\hat X$~\cite{CohenBook}. This induces a quantum noise onto the classically deterministic quantity. Arguably, the most famous consequence of this variance is the Heisenberg uncertainty relation, which shatters the classical view of a fully deterministic universe. A similarly fundamental aspect is given by single-photon sources, where no more than one photon can be emitted or absorbed at the same time. Field correlations that include two or more simultanuous excitations or deexcitations vanish identically. The fluorescent emission from these systems is in turn nonclassical, attaining statistical properties that are incompatible with solutions of the classical Maxwell equations. 

One of the most famous and measured correlation functions is the second-order correlation function $g^{(2)}(\tau)$, first applied by Hanbury-Brown and Twiss~\cite{HBT1956} in the fifties of the last century. In their original proposal, they looked at the classical version of this correlation function to measure the size of distant stars. Following the pioneering works of Sudarshan~\cite{Sudarshan63} and Glauber~\cite{Glauber63}, the Hanbury-Brown and Twiss measurement setup became a cornerstone to reveal quantum features of light, in particular antibunching~\cite{Kimble77}, and sub-Poissonian photon statistics~\cite{Short83}. Beyond this quantum application, a spatial analysis $g^{(2)}(d)$ of two Rydberg excitons with distance $d$ was recently used to visualize Rydberg blockade~\cite{Valentin2018}. In solid-state optics, $g^{(2)}(0)$ of a single-mode emission field is used to evaluate the single-photon character of the source field~\cite{Michler2000,Vuckevic2012}. If this value falls below $1/2$ the light source is considered a good single-photon source. Some limitations of this criterion and the proposal for using higher-order correlations has been brought up multiple times~\cite{Rundquist2014,Laussy16}. It was also shown that for sub-Poissonian light the average photon number is limited with clear hard boundaries~\cite{Zubizarreta2017}.
In a recent work~\cite{g2paper}, we analyzed the information that can be gained from $g^{(2)}(0)<1/2$ concerning the single-photon projection of the underlying quantum state. While this criterion at face value is limited to sub-Poissonian light, using additional information necessary to evaluate the actual projection on the single-photon Fock state allows to quantify some classical light fields as well. In short, the second-order correlation function has become a major resource for information in a plethora of different applications in modern classical and quantum physics.

In contrast, the higher-order correlation functions $g^{(k)}$, first introduced by Glauber~\cite{Glauber63}, have been shunned for many years. This is in part due to the complicated process of measuring this function. Only a decade ago, experimental accessibility became possible for significantly larger orders~\cite{Silberhorn2010}, thanks to a theoretical proposal developed a few years prior~\cite{Shchukin2006}. Nevertheless, recently, higher-order correlation functions have become relevant for different quantum systems such as optomechanical setups~\cite{Mukherjee2019}, photon-added and subtracted squeezed coherent states~\cite{Thapliyal2017}, and noisy twin beams~\cite{Arkhipov2018}. From an experimental point of view, twin beams combined with post-selection were used to anlyze the quality of the obtained correlation functions for different measurement quantities such as the field intensities or the explicite photon numbers~\cite{Perina2017,Perina2019}.
In all of these applications higher-order correlation functions were employed exclusively to show the basic nonclassicality feature of higher-order sub-Poissonian photon statistics~\cite{Pathak2006}. To this day, this phenomenon, also referred to as higher-order photon blockade, is intrinisically linked to quantum effects~\cite{Miranowicz19}. Another recent application, identifying entangled bunches of photons in the emission of two-level arrays~\cite{Liberal19}, also is intimately connected to quantum states without classical analogue. We are still far from the versatility known from $g^{(2)}$ and different applications of these functions not limited to nonclassical quantum states are of fundamental interest. 

The aim of this work is to generalize~\cite{g2paper} to higher-order correlation functions $k>2$ and thus provide such a novel set of information gained from $g^{(k)}(0)$. The main focus is on deriving generalized formulas for the following results from the case $k=2$: (a) when $g^{(k)}(0)$ falls below the value attained for the Fock state $|k\rangle$, there is a nonzero lower bound for the projection of the state  onto the subspace spanned by the Fock states with photon number less than $k$. We will refer to this subspace as the sub-$k$ space from now on. (b) A nonzero vacuum projection artificially enhances $g^{(k)}(0)$ for a state with otherwise fixed ratios of Fock-state projections. We derive an effective correlation function $\tilde g^{(k)}(0)$ to account for these vacuum effects. (c) With $\tilde g^{(k)}(0)$, we are able to determine a lower bound for the ratio of one-to-$(k-1)$ Fock state projections relative to $k$-or-more Fock state projections. (d) The effective correlation function also yields information for some classical states of light. This shows that the criteria, while at face value implying $k$th-order sub-poissonian fields, are actually independent of nonclassicality conditions. (e) It is possible to obtain $\tilde g^{(k)}(0)$ directly by combining balanced homodyne correlation measurements with post-selection.
Beyond the generalization from $k=2$ we also present a large-$k$ approximation, which serves as a valid lower bound for all $k$.

The paper is organized as follows. In Sec.~\ref{sec.Not,Rev}, we provide the notation used throughout this work and give a brief summary of the major results of~\cite{g2paper}. Followed by that, we give the generalized proof that having $g^{(k)}(0)$ lower than a specific minimum guarantees a non-zero projection onto the sub-$k$ space in Sec.~\ref{sec.proof}. In Sec.~\ref{sec.amplitudes} we give lower bounds for both the absolute projection onto the sub-$k$ space, as well as the projection of the sub space spanned by Fock states  from 1 to $k-1$ photons (the sub-$\tilde k$ space) relative to the sub space from $k$ to infinity (the super-$k$ space). Each of these results is a generalization of the previous special analysis for $k=2$, meaning there are explicite states, for which the bounds are reached. Then in Sec.~\ref{sec.large-k} we compute an analytical large-$k$ approximation of the bounds. All these results will be applied to known states in Sec.~\ref{sec.applications}. Sec.~\ref{sec.meas} is dedicated to a measurement scheme for the effective $k$th-order correlation function. Finally, we give conclusions in Sec.~\ref{sec.concl}.

\section{Notation and case $k=2$}\label{sec.Not,Rev}
The general form of a $k$-th order correlation function is given by
\begin{equation}
	g^{(k)}(0)=\frac{\langle(\hat E^{(-)})^k(\hat E^{(+)})^k\rangle}{\langle\hat E^{(-)}\hat E^{(+)}\rangle^k},\ k\in\mathbb N\geq2,
\end{equation}
where $\hat E^{(+)}$ and $\hat E^{(-)}=[\hat E^{(+)}]^\dagger$ are the positive and negative frequency field amplitudes, usually evaluated in the steady state of the system, respectively. Obviously, for no projection on $k$ or more photons, $g^{(k)}(0)=0$, but in general already
\begin{equation}
	g^{(k)}(0)<1\label{eq.genNC}
\end{equation}
proves nonclassicality of the underlying quantum state of light~\cite{Pathak2006}, called $k$th-order sub-Poissonian light. Moreover, as all operators are normally-ordered in $g^{(k)}(0)$, one can connect these field correlation functions to the source fields emitted from their origin and in turn to the system operators (usually atomic or atom-like) of that source~\cite{WelVo}. Likewise, as the intensity is scaled out in this function, for a single-mode field we can write $g^{(k)}(0)$ only in terms of creation(annihilation) operators $\hat a^\dagger(\hat a)$ as
\begin{equation}
	g^{(k)}(0)=\frac{\langle\hat a^{\dagger k}\hat a^k\rangle}{\langle\hat a^\dagger\hat a\rangle^k}.
\end{equation}
We will analyze this function throughout the manuscript.

For the sake of clarity and brevity, we introduce the following notation to be used from now on. The order $k$ of the correlation function will be arbitrary but fixed, unless otherwise stated; the index $k$ is thus always meant to represent the $k$th-order correlation function $g^{(k)}(0)$. We will only consider the correlation function at time delay zero, hence $g^{(k)}=g^{(k)}(0)$ for any state. Furthermore, when explicitely calculating $g^{(k)}$ for a given state $\hat\varrho$, we use the form $g^{(k)}[\hat\varrho]$ or $g^{(k)}[|\psi\rangle]$ for a pure state $|\psi\rangle$. The Fock states are denoted $|n\rangle$, $n\in\mathbb N$ and the photon statistics are $p_n=\langle n|\hat\varrho|n\rangle$. For later purposes we define $g^{(k)}_\text{min}=g^{(k)}[|k\rangle]$.

The photon statstics are split into projections onto the sub-$k$ and super-$k$ space as defined in Sec.~\ref{sec.intro}, rationalizing the introduction of the shorthand
\begin{equation}
	P=\sum\limits_{n=0}^{k-1} p_n, \quad Q=\sum\limits_{n=k}^\infty p_n=1-P. 
\end{equation}
Furthermore, as the vacuum contributions will become relevant, we also use $\tilde P=P-p_0$. For the sake of avoiding pathologies, we will always assume to have states with $P,Q\neq0$.
With the split of the Hilbert space into these two subspaces, we introduce corresponding states 
\begin{equation}
	\hat\varrho_{P}=\frac{1}{P}\sum\limits_{n=0}^{k-1}p_n|n\rangle\langle n|,\quad \hat\varrho_{Q}=\frac{1}{Q}\sum\limits_{n=k}^{\infty}p_n|n\rangle\langle n|, 
\end{equation}
as well as their average photon number $N_{P(Q)}=\text{Tr}\{\hat n\hat\varrho_{P(Q)}\}$ with the obvious condition $N_P\leq k-1$, $N_Q\geq k$. Note that, as in the case for $k=2$, all the information gathered from measuring $g^{(k)}$ is contained within the $p_n$, and thus, we can use $\hat\varrho_{P(Q)}$ for a general description of arbitrary quantum states. In the same way as $N_{P(Q)}$, we define $k$th-order correlation functions $g^{(k)}_{P}=g^{(k)}[\hat\varrho_P]=0$ and $g^{(k)}_{Q}=g^{(k)}[\hat\varrho_Q]>0$. 

With the above introduced notation, let us shortly review the main results and steps taken within~\cite{g2paper}, i.e., the case $k=2$. Starting from the well-known result for Fock states having the property $g^{(2)}[|n\rangle]=1-1/n$, $n\geq 1$, we showed that $g^{(2)}$ is quasiconcave (but not quasiconvex),
\begin{align}
	g^{(2)}[s\hat\varrho_1+(1-s)\hat\varrho_2]\geq& \text{min}\{g^{(2)}[\hat\varrho_1],g^{(2)}[\hat\varrho_2]\}\label{eq.g2superpos}
\end{align}
for arbitrary quantum states $\hat\varrho_{1,2}$ and $s\in[0,1]$.
This yielded in general the statement that for
\begin{equation}
	g^{(2)}<1/2=g^{(2)}[|2\rangle]=g^{(2)}_\text{min},
\end{equation}
we have $\tilde P=p_1>0$. The absolute amplitude of $\tilde P$ does not follow from $g^{(2)}$ alone, but the relative amplitude
\begin{align}
	\frac{\tilde P}{Q}\geq&\frac{2\sqrt{1-2\tilde g^{(2)}}}{1-\sqrt{1-2\tilde g^{(2)}}},\label{eq.g2result}\\
	\tilde g^{(2)}=&(1-p_0)g^{(2)}.
\end{align}
The only variable on the right-hand side of Eq.~(\ref{eq.g2result}), $\tilde g^{(2)}$, is called the effective second-order correlation function. The scaling incorporates the effects of the vacuum contribution $p_0$, thus generating a vacuum-independent lower bound for $\tilde P/Q$. In case we have no information on vacuum we must assume $p_0=0$. Equality of (\ref{eq.g2result}) is given, if no more than two-photon projections are present, i.e. $Q=p_2$. The result can also be given as a lower bound for the sum of vacuum and single-photon projection, which is $P$ for $k=2$, and reads as
\begin{align}
		P\geq&\frac{2\sqrt{1-2\tilde g^{(2)}}}{1+\sqrt{1-2\tilde g^{(2)}}}.\label{eq.newIP}
\end{align}
Finally, we noted that as weakly excited states have large vacuum contributions $p_0$, we can also analyze coherent and thermal states in this regime, showing the independence of the original criterion from the sub-Poissonian light condition. 

It has to be stated in this context that large vacuum contributions are not a goal in single-photon research. Quite the opposite, they not only cover potential single-photon projections. In experiments the corresponding low signal-intensity also diminishes the signal-to-noise ratio rendering quantitative analysis impossible. Due to this problem, a scheme was proposed~\cite{Lachman16} and later realized~\cite{Moreva17} to detect nonclassicality, using click-detectors and building correlations only from large vacuum contributions, i.e., from the condition of no click in the detctors. In contrast, we used the additional information given by $p_0$ to evaluate the actual value of the single-photon projection $p_1$, which was impossible from just $g^{(2)}$.

\section{Nonzero Projection on sub-$k$ space}\label{sec.proof}
We proof the nonzero projection on the sub-$k$ space in a two-step process. In the first step we show that for Fock states $g^{(k)}$ is monotone increasing with the photon number, i.e. $g^{(k)}[|n\rangle]\leq g^{(k)}[|n+1\rangle]$. 
The $k$th-order correlation function for Fock states reads
\begin{equation}
	\begin{split}
		g^{(k)}[|n\rangle]=&\frac{1}{n^k}\frac{n!}{(n-k)!},\quad n\geq k,\\
		g^{(k)}[|n\rangle]=&0,\quad n< k.
	\end{split}
\end{equation}
Consider the ratio of $g^{(k)}$ for consecutive Fock states
\begin{equation}
	\frac{g^{(k)}[|n\rangle]}{g^{(k)}[|n+1\rangle]}=\left(1+\frac{1}{n}\right)^k\left(1-\frac{k}{n+1}\right).
\end{equation}
This positive function should remain lower or equal to 1 for all combinations $2\leq k\leq n$. Obviously, for $n\to\infty$, this ratio becomes one.
Let us for the moment extend the range of $n$ to real numbers larger or equal to a fixed $k$. In that case the derivative with respect to $n$ reads as
\begin{align}
	\frac{d}{dn}\left[\frac{g^{(k)}[|n\rangle]}{g^{(k)}[|n+1\rangle]}\right]=&\left(1+\frac{1}{n}\right)^{k-1}\frac{-k}{n^2}\left(1-\frac{k}{n+1}\right)\nonumber\\
	&+\left(1+\frac{1}{n}\right)^k\frac{k}{(n+1)^2}\\
	=&\frac{k(k-1)}{n^2(n+1)}\left(1+\frac{1}{n}\right)^{k-1}>0.
\end{align}
The function is thus positive, always increasing with $n$ and goes to 1 for $n\to\infty$, from which we can conclude that $g^{(k)}[|n\rangle]$ is monotone increasing.

In the second step we make use of the ability to have a unified treatment for coherent and incoherent superpositions as all expectation values in our calculation only concern diagonal entries on the density matrix when written in Fock-state basis, cf. the argument for $k=2$ in~\cite{g2paper}. Hence, we only need to show that $g^{(k)}$ is quasiconcave, i.e.,
\begin{equation}
	g^{(k)}[s\hat\varrho_1+(1-s)\hat\varrho_2]\geq \text{min}\{g^{(k)}[\hat\varrho_1],g^{(k)}[\hat\varrho_2]\}\label{eq.gksuperpos}
\end{equation}
for every $\hat\varrho_{1,2},s\in[0,1]$.
Denoting for the two states $g_i=g^{(k)}[\hat\varrho_i]$ and $n_i=\textrm{Tr}\{\hat\varrho_i\hat a^\dagger\hat a\}$ with $i=1,2$ we find
\begin{equation}  
\begin{split}
g^{(k)}[\hat\varrho=s\hat\varrho_1+(1-s)\hat\varrho_2]=&\frac{sn^k_1g_1+(1-s)n_2^kg_2}{[sn_1+(1-s)n_2]^k}.\label{eq.g(k)[rho]}
\end{split}
\end{equation}
Without loss of generality, we can set $r=n_2/n_1\geq0$, and $g_2=tg_1$, $t\in[0,1]$ and rewrite the formula as
\begin{equation}  
	g^{(k)}[\hat\varrho]=g_1\frac{s+(1-s)r^kt}{[s+(1-s)r]^k}.
\end{equation}
Varying $s$ from 0 to 1, $g^{(k)}$ shifts from $g_2$ to $g_1$, i.e., it does not decrease overall. 
The derivative with respect to $s$ reads as
\begin{equation}
	\frac{d}{ds}g^{(k)}=g_1\frac{[1-r^kt][s+(1-s)r]-k(1-r)[s+(1-s)r^kt]}{[s+(1-s)r]^{k+1}}.
\end{equation}
It has a positive denominator and a numerator linear in $s$, indicating no more than one extreme point. Consequently, in order to not be quasiconcave $g^{(k)}$ needs to be decreasing at the beginning, that is 
\begin{equation}
\left.\frac{d}{ds}g^{(k)}\right|_{s=0}=\frac{g_1}{r^k}\left[1+(k-1)r^kt-kr^{k-1}t\right]\leq0.\label{eq.numer}
\end{equation}
Possible negativities depend on the roots of the square bracket in Eq.~(\ref{eq.numer}), which can be rewritten as
\begin{equation}
	1-tr^k(1-k+\tfrac{k}{r}),
\end{equation}
yielding as condition for a decreasing slope
\begin{equation}
 1\leq\frac{1}{t}\leq r^k(1-k+\tfrac{k}{r}).\label{eq.numer3}
\end{equation}
As $k\geq2$, the right-hand side of Eq.~(\ref{eq.numer3}) is only positive in the interval $0\leq r\leq k/(k-1)$, and zero at its boundaries. The maximum in between is at $r=1$ yielding $t=1$ as the only solution, where $g^{(k)}$ does not increase at $s=0$. For this case with $n_1=n_2$ and $g_1=g_2$, $g^{(k)}[\varrho]$ is constant, as it can not distinguish between the two states, and thus also does not decrease. 
Thus, for all cases $g^{(k)}[\varrho]$ is quasiconcave and there is a nonzero projection on the sub-$k$ Fock space if 
\begin{equation}
	g^{(k)}<g^{(k)}_\text{min}=\frac{k!}{k^k}.
\end{equation} 
Note that we can also conclude that $g^{(k)}_Q\geq g^{(k)}[|k\rangle]=g^{(k)}_\text{min}$.

\begin{figure}[ht]
	\includegraphics[width=8.6cm]{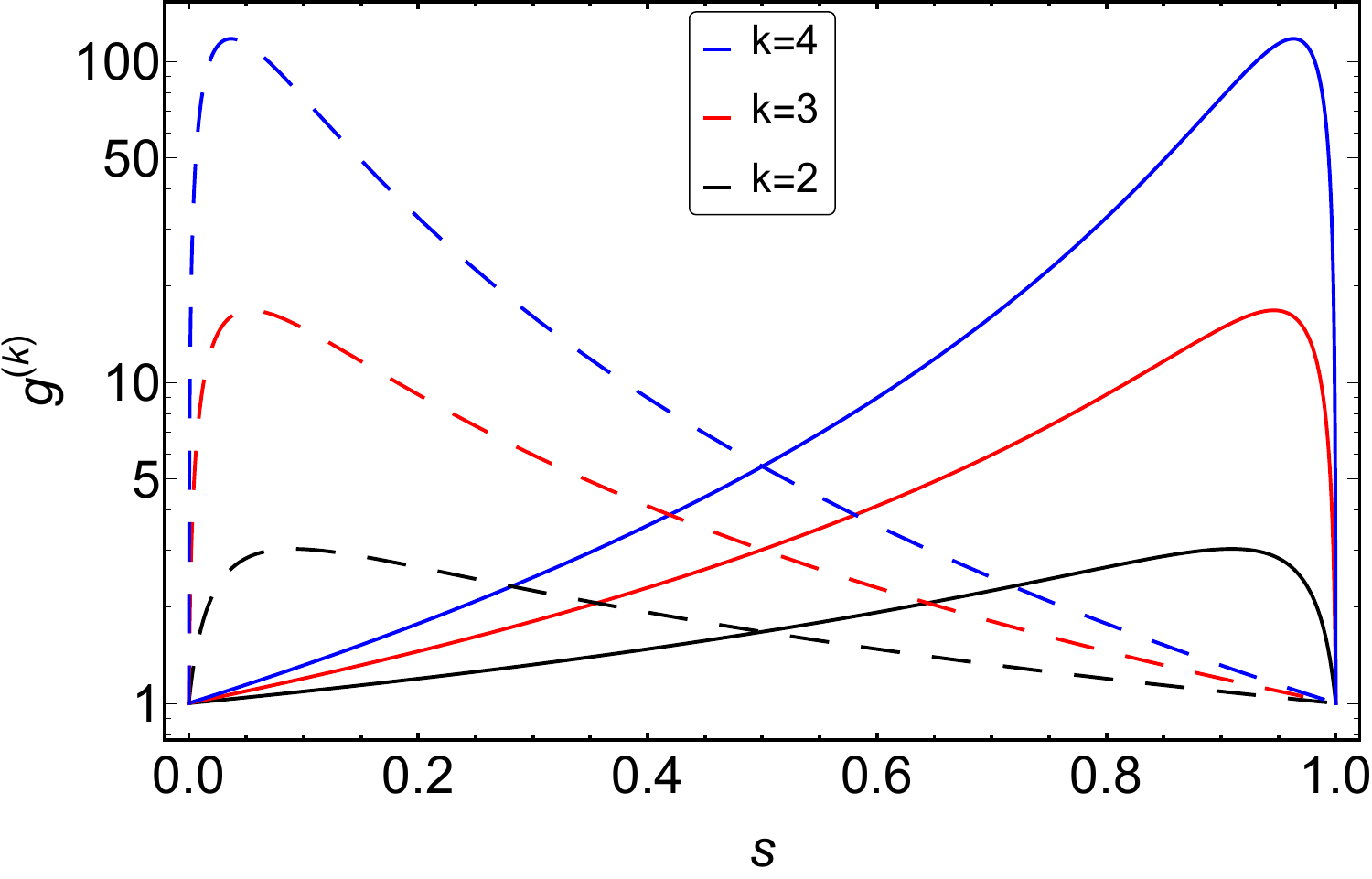}
	\caption{$k$th-order correlation function for the incoherent mixing of two coherent states ($g_i=1$) with $r=10$ (solid lines), and $r=1/10$ (dashed lines). From top to bottom, each pair of inverted lines represents $k=4,3,2$.}\label{fig.cohsup}
\end{figure}

A few comments are in order. While we have technically only shown that $P\neq0$ so far, in comparison to $\tilde P$ in~\cite{g2paper}, the extension to this case follows simply from setting $\varrho_2=|0\rangle\langle0|$ in Eq.~(\ref{eq.g(k)[rho]}), which leads to 
\begin{equation}  
g^{(k)}[\hat\varrho=s\hat\varrho_1+(1-s)|0\rangle\langle0|]=\frac{g_1}{s^{k-1}}.
\end{equation}
Vacuum itself only increases the value of $g^{(k)}$, whereas the lowering below $g^{(k)}_\text{min}$ requires a nonzero projection on a Fock state between $1$ and $k-1$. 
We observe that $g^{(k)}$ is not quasiconvex, as there is no general upper bound to $g^{(k)}[\hat\varrho]$. For two states $\hat\varrho_i$ with equal $g_i$ $(t=1)$, but different average photon numbers $(r\neq1)$, $g^{(k)}$, the superposition has a maximum value of
\begin{equation}
	g^{(k)}=g_1\frac{1}{r^{k-1}(r-1)}\frac{(r^k-1)^k}{k^k}\frac{(k-1)^{k-1}}{(r^{k-1}-1)^{k-1}}.
\end{equation}
One can easily deduce for $r\gg1$ that
\begin{equation}
	g^{(k)}\approx g_1\frac{(k-1)^{k-1}}{k^k}r^{k-1},
\end{equation}
and, correspondingly, for $r\ll1$ there is a limit with $r^{-(k-1)}$. As an example, we plotted the result for the incoherent mixing of two coherent states in Fig.~\ref{fig.cohsup}.

Finally, one corollary should be mentioned. As we have shown the general quasiconcave property of $g^{(k)}$ and the monotonicity of $g^{(k)}[|n\rangle]$, we can also generalize the lower bound argument to any $n\geq k$. That means, whenever $g^{(k)}<g^{(k)}[|n\rangle]$ with $n\geq k$, a nonzero projection of the sub-$n$ space exists. As
\begin{equation}
	\lim\limits_{n\to\infty}g^{(k)}[|n\rangle]=1,
\end{equation}
we conclude that for any state with $g^{(k)}<1$, that is for all states for which $k$th-order sub-Poissonian statistics are found, there exists a number $n\geq k$ with a nonzero projection on the sub-$n$ space. All subsequent results can be modified for this generalized result, but for the sake of brevity and clarity we stick to the case $n=k$. This is in direct connection to the results of~\cite{Zubizarreta2017}, in which the authors analyzed the relation between a low $g^{(2)}$ and the average photon number of the underlying quantum state. As a major result, it was shown that for sub-Poissonian light ($g^{(2)}<1)$, there exists an upper bound on the average photon number, as well as upper bounds on the $p_n$ ($n\geq2$) above a certain threshold. Hence, when the maximum projection $Q$ for all the states above $n$ falls below one, there must be a nonzero projection $P$. In short this means that sub-Poissonian light (to any order $k$), originally connected to low variance of photon statistics, also implies a limit on the average photon number, see also the very recent work on sub-Poissonian fields in microlasers~\cite{Ann19}. 

\section{Lower bounds of $P$ and $\tilde P/Q$}\label{sec.amplitudes}
With the knowledge of the existence of a nonzero $P$, this section is aimed at deriving different bounds on the amplitude of the sub-$k$ projection.
Splitting $g^{(k)}$ into two sums at $k$, we obtain
\begin{align}
	g^{(k)}=&\frac{\sum\limits_{n=k}^\infty n^kp_ng^{(k)}[|n\rangle]}{\left[\sum\limits_{n=0}^{k-1} np_n+\sum\limits_{n=k}^\infty np_n\right]^k}\\
	\sum\limits_{n=0}^{k-1} np_n=&\sqrt[k]{\frac{1}{g^{(k)}}\sum\limits_{n=k}^\infty n^kp_ng^{(k)}[|n\rangle]}-\sum\limits_{n=k}^\infty np_n.
\end{align}
In terms of the above defined states $\hat \varrho_{P(Q)}$ this is equivalent to
\begin{align}
	N_PP=&N_Q\left[\sqrt[k]{\frac{g^{(k)}_Q}{g^{(k)}}Q}-Q\right]\label{eq.exactfinal}
\end{align}
So far this equation is exact. It connects the projection on the sub-$k$ space, namely $P$, to the the projection on the super-$k$ space, namely $Q$. 
Applying the monotonicity of $g^{(k)}[|n\rangle]$ and the average photon number of Fock states in order to get a lower bound on $P$ yields
\begin{equation}
	P\geq\frac{k}{k-1}\left[\sqrt[k]{\frac{g^{(k)}_\text{min}}{g^{(k)}}Q}-Q\right].\label{eq.PtoQ}
\end{equation}
It should be noted, that these inequalities have tight bounds. They become equations for the only nonzero projections being on the Fock states $|k-1\rangle$ and $|k\rangle$.

\begin{figure}[ht]
\includegraphics[width=8.6cm]{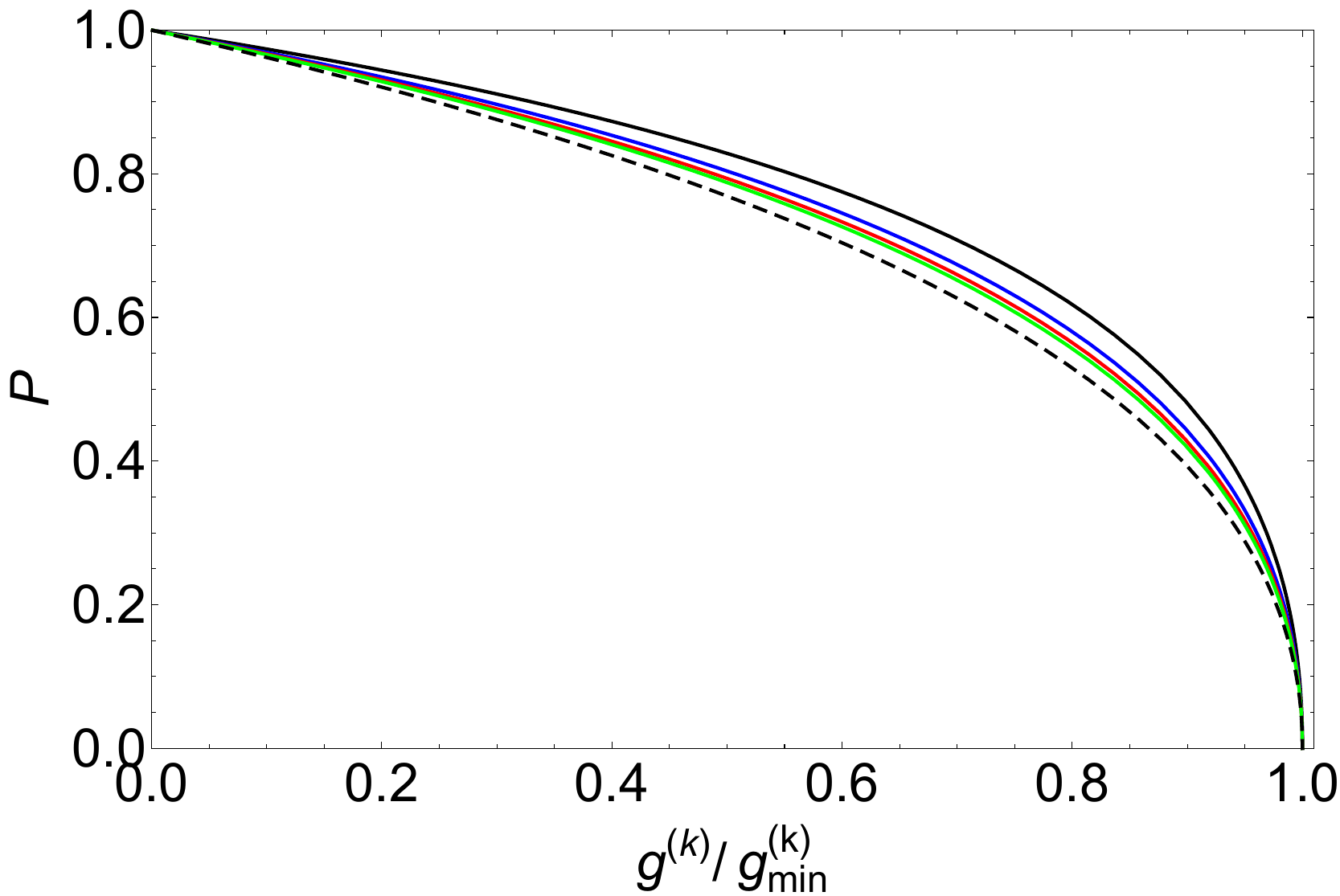}
\caption{The probability $P$ of sub-$k$ photon numbers as a function of $g^{(k)}/g^{(k)}_\text{min}\leq1$. From top to bottom the solid functions represent $k=2,3,4,5$. The dashed line represents $k=100$ showing the large-$k$ asymptotics.}\label{fig.probab1}
\end{figure}

With $P=1-Q$, $Q$  is the only unknown quantity in Eq.~(\ref{eq.PtoQ}). We know $Q\in]0,1[$ and one can easily prove that due to the monotonicity of the terms in Eq.~(\ref{eq.PtoQ}) with respect to $Q$ only one solution $Q_\text{max}$ exists. For $k=2$,  $Q_\text{max}$ was determined analytically. In the general case it can be computed numerically. For $P$ it follows the solution
\begin{equation}
	P\geq P_\text{min}=\frac{k}{k-1}\left[\sqrt[k]{\frac{g^{(k)}_\text{min}}{g^{(k)}}Q_\text{max}}-Q_\text{max}\right],\label{eq.PtoQ2}
\end{equation}
with the minimal sub-$k$ space projection $P_\text{min}=1-Q_\text{max}$.
This equation is a generalized version of Eq.~(\ref{eq.newIP}) for arbitrary $k$. It states that for $g^{(k)}<g^{(k)}_\text{min}$ the projection on the sub-$k$ space has a non-zero lower bound. 
We have visualized $P$ for different $k$ in Fig.~\ref{fig.probab1}. One can see that the probabilities are smooth functions of the ratio $g^{(k)}/g^{(k)}_\text{min}$ and decreasing for increasing $k$. Moreover, the functions appear to stabilize for large $k$, indicating the existence of a general lower bound to be determined later. It should be noted that the difference between the low-$k$ and large-$k$ boundaries is very small, the maximum deviation between the probability $P$ for $k=2$ and $k=100$ is 0.09.

Similar to~\cite{g2paper}, we can also easily determine an analytic approximation for low $g^{(k)}\ll g^{(k)}_\text{min}$. Assuming in Eq.~(\ref{eq.PtoQ2}) that $Q_\text{max}\ll1$, we obtain
\begin{equation}
	P_\text{min}\approx 1-\frac{(k-1)^k}{k!}g^{(k)}.
\end{equation}
Expanding Eq.~(\ref{eq.newIP}) for low $\tilde g^{(2)}$ gives the exact same formula except for the effective correlation function, indicating that we so far avoided discussing the effect of vacuum.

In order to better understand on the influence of vacuum, we first note that on the left-hand side of Eq.~(\ref{eq.exactfinal}), the vacuum term $p_0$ does not contribute directly, as the average photon number was calculated and  $N_PP\leq(k-1)\tilde P$. Thus, we can write
\begin{equation}
	\tilde P\geq\frac{k}{k-1}\left[\sqrt[k]{\frac{g^{(k)}_\text{min}}{g^{(k)}}Q}-Q\right]\label{eq.tildePtoQ}
\end{equation}
with $\tilde P=P-p_0$, connecting the sub-$\tilde k$ space, spanned by the Fock states of photon number $1$ to $k-1$, to the super-$k$ space of $Q$. However, $p_0$ becomes an additional free parameter that shifts both $Q_\text{max}$ and $\tilde P$ down making their absolute amplitudes indeterminable from only $g^{(k)}$. For $k=2$, solving this equation for $\tilde P/Q$ yielded automatically the effective second-order correlation function $\tilde g^{(2)}=(1-p_0)g^{(2)}$. The origin of this term only became clear in hindsight as compensating the effect of vacuum, which itself enhances $g^{(2)}$ for fixed $\tilde P/Q$. In order to generalize the influence of vacuum to higher $k$, we will give its physical explanation first. Consider a state $\hat\varrho_0$ with no vacuum $(p_0=0)$ and a given ratio $\tilde P/Q>0$. Now we can include vacuum in the state as
\begin{equation}
	\hat\varrho_{p_0}=p_0|0\rangle\langle 0|+(1-p_0)\hat\varrho_0.
\end{equation}
The ratio $\tilde P/Q$ stays fixed, but $g^{(k)}$ gets scaled as
\begin{equation}
	g^{(k)}[\hat\varrho_{p_0}]=\frac{(1-p_0)\text{Tr}\{\hat\varrho_{0}\hat a^{\dagger k}\hat a^k\}}{[(1-p_0)\text{Tr}\{\hat\varrho_0\hat a^\dagger\hat a\}]^k}=\frac{g^{(k)}[\hat\varrho_0]}{(1-p_0)^{k-1}}.
\end{equation}
Vacuum artificially enhances the value of $g^{(k)}$, motivating the definition of an effective $k$th-order correlation function
\begin{equation}
	\tilde g^{(k)}=(1-p_0)^{k-1}g^{(k)},
\end{equation}
which in turn gives a vacuum-independent assessment of the sub-$k$ and sub-$\tilde k$ spaces.

\begin{figure}[ht]
\includegraphics[width=8.6cm]{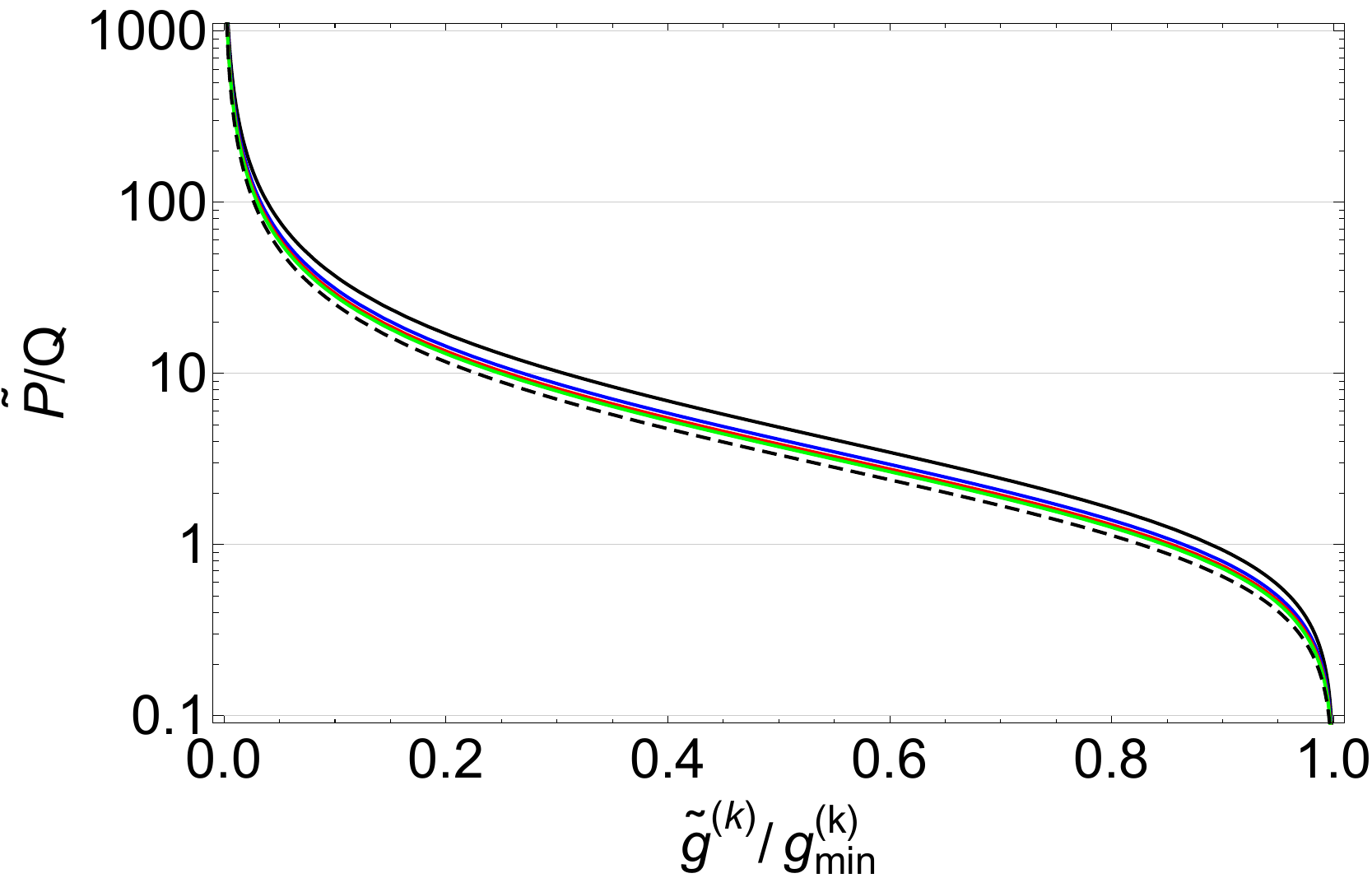}
\caption{Relative probability $\tilde P/Q$ as a function of $\tilde g^{(k)}/g^{(k)}_\text{min}\leq1$. The curves are the same as in Fig.~\ref{fig.probab1}.}\label{fig.probrel1}
\end{figure}

With the knowledge of the effective $k$th-order correlation function in mind let us return to Eq.~(\ref{eq.PtoQ}) and its solution $Q_\text{max}$. If we write out $Q$ on the left-hand side of Eq.~(\ref{eq.tildePtoQ}) and define $Q=(1-p_0)\tilde Q$, we obtain
\begin{align}
	1-\tilde Q\geq&\frac{k}{k-1}\left[\sqrt[k]{\frac{g^{(k)}_\text{min}}{\tilde g^{(k)}}\tilde Q}-\tilde Q\right].
\end{align}
The result is structually identical to Eq.~(\ref{eq.PtoQ}), just for $\tilde Q$ and $\tilde g^{(k)}$. That means, $\tilde Q$ has the same solution as $Q$, but for $\tilde g^{(k)}$ instead of $g^{(k)}$, yielding
\begin{equation}
	Q_\text{max}[p_0,g^{(k)}]=(1-p_0)Q_\text{max}[0,\tilde g^{(k)}].
\end{equation}
Note that the case $p_0=0\leftrightarrow\tilde g^{(k)}=g^{(k)}$ is included in this generalization. Furthermore, inserting this solution into Eq.~(\ref{eq.tildePtoQ}), we find
\begin{align}
	\frac{\tilde P}{Q}\geq&\frac{k}{k-1}\left[\sqrt[k]{\frac{g^{(k)}_\text{min}}{g^{(k)}}\frac{1}{Q^{k-1}_\text{max}[p_0,g^{(k)}]}}-1\right]\\
	=&\frac{k}{k-1}\left[\sqrt[k]{\frac{g^{(k)}_\text{min}}{\tilde g^{(k)}}\frac{1}{Q^{k-1}_\text{max}[0,\tilde g^{(k)}]}}-1\right].\label{eq.ratioOpt}
\end{align}
The right-hand side of Eq.~(\ref{eq.ratioOpt}) does not contain $p_0$ or $g^{(k)}$ individually, but only $\tilde g^{(k)}$. Hence, we have proven that the relevant quantity for the lower bound of $\tilde P/Q$ is the effective $k$th-order correlation function $\tilde g{(k)}$, in accordance with the main result of~\cite{g2paper}. Again, we plot the results in Fig.~\ref{fig.probrel1} for the same cases as in Fig.~\ref{fig.probab1}. In the logarithmic scaling the variation with $k$ appears even less significant, emphasizing the necessity to consider the large-$k$ approximation.

Two important conclusions can be drawn. First, we can also use Eq.~(\ref{eq.ratioOpt}) to further optimize the lower bound of $P$. Therefore we use the exact same argument as in Eqs.~(\ref{eq.g2result}-\ref{eq.newIP}), now with the right-hand-side of Eq.~(\ref{eq.ratioOpt}), yielding
\begin{align}
	P\geq&\frac{\sqrt[k]{\frac{g^{(k)}_\text{min}}{\tilde g^{(k)}}\frac{1}{Q^{k-1}_\text{max}[0,\tilde g^{(k)}]}}-1}{\sqrt[k]{\frac{g^{(k)}_\text{min}}{\tilde g^{(k)}}\frac{1}{Q^{k-1}_\text{max}[0,\tilde g^{(k)}]}}-\frac{1}{k}}\\
	&=\frac{kP_\text{min}[0,\tilde g^{(k)}]}{(k-1)+P_\text{min}[0,\tilde g^{(k)}]}.\label{eq.Pfull}
\end{align}
Herein $P_\text{min}[0,\tilde g^{(k)}]=1-Q_\text{max}[0,\tilde g^{(k)}]$. Note that while Eq.~(\ref{eq.Pfull}) gives a larger lower bound than Eq.~(\ref{eq.PtoQ2}), its effect is negligible for $\tilde g^{(k)}\ll g^{(k)}_\text{min}$ or $k\gg1$.

Second, even with the effective $k$th-order correlation function, we can only determine a lower bound $\tilde P/Q$, not $\tilde P$, consistent with the observation in~\cite{g2paper} that the single-photon projection itself requires additional information. However, with explicite knowledge of the vacuum projection $p_0$ and Eq.~(\ref{eq.Pfull}), absolute lower and upper bounds can easily be established as
\begin{equation}
	1-p_0 \geq \tilde P \geq \frac{kP_\text{min}[0,\tilde g^{(k)}]}{(k-1)+P_\text{min}[0,\tilde g^{(k)}]}-p_0.
\end{equation}
As a consequence of this necessary addition, we will see in Sec.~\ref{sec.applications}, that determining the amplitude $\tilde P$ from $g^{(k)}$ constitute an application of the higher-order correlation function, which is not bound by nonclassical quantum states.

\section{Large-$k$ approximation}\label{sec.large-k}
As can be seen from the dashed curves in Figs.~\ref{fig.probab1},\ref{fig.probrel1}, for large $k$ the probability $P$ and the relative amplitude $\tilde P/Q$ stabilize at a smooth function. This function serves as a general lower bound, depending only on the ratio $\tilde g^{(k)}/g^{(k)}_\text{min}$, which from now on, we will denote as $R$ with $0\leq R\leq1$. In order to analyze this case, let us first turn back to Eq.~(\ref{eq.PtoQ}) in the form
\begin{equation}
	1-Q_\text{max}=\frac{k}{k-1}\left(\sqrt[k]{\frac{Q_\text{max}}{R}}-Q_\text{max}\right).\label{eq.appStart}
\end{equation}
As $k/(k-1)>1$, the root must be smaller or equal to one, with equality only given for $Q_\text{max}=R=1$. Thus, we can rewrite the root and make a series expansion as
\begin{align}
\sqrt[k]{\frac{Q_\text{max}}{R}}=&\sqrt[k]{1- \left(1-\frac{Q_\text{max}}{R}\right)}=\sqrt[k]{1- x}\\
				=&1-\frac{1}{k}x-\frac{1}{k}\left(1-\frac{1}{k}\right)\frac{x^2}{2}-\dots.
\end{align}
For large $k$ the term $1/k$ in parentheses of the form $n-1/k$, $n\in\mathbb N$ can be neglected, leaving the Taylor expansion of the natural logarithm as
\begin{align}
\sqrt[k]{\frac{Q_\text{max}}{R}}\approx&1+\frac{1}{k}\left[-x-\frac{x^2}{2}-\frac{x^3}{3}-\dots\right]\\
											=&1+\frac{1}{k}\ln(1-x)=1+\frac{1}{k}\ln\left(\frac{Q_\text{max}}{R}\right).
\end{align}
Inserting this result back into Eq.~(\ref{eq.appStart}), the explicit $k$-dependencies cancel yielding
\begin{align}
1-Q_\text{max}=&\ln\left(\frac{R}{Q_\text{max}}\right),\ \text{or}\ P_\text{min}=\ln\left(\frac{R}{1-P_\text{min}}\right).
\end{align}
Thus, we have a large-$k$ behaviour, where the only $k$-dependence is given via $g^{(k)}_\text{min}$ in $R$, yielding a general lower bound of $P_\text{min}(R)$. To formulate this implicite solution with explicite functions, we calculate the derivative $P'_\text{min}(R)$ to obtain
\begin{equation}
	P'_\text{min}(R)=-\frac{\exp[-P_\text{min}(R)]}{P_\text{min}(R)}.
\end{equation}
Including the boundary condition $P_\text{min}(0)=1$ this differential equation has the unique solution
\begin{equation}
	P_\text{min}(R)=1+W_0\left(-\frac{R}{e}\right)\label{eq.largeP}
\end{equation}
with $W_0(x)$ the Lambert-$W$ function with the upper branch for $x\in[-1/e,0]$. Finally, the large-$k$ approximation for the relative amplitude follows as $P_\text{min}/(1-P_\text{min})$.

\section{Application}\label{sec.applications}
The inclusion of vacuum effects allows us to describe the projection on the sub-$k$ space not just for the case of sub-Poissonian light. 
To illuminate this thought consider a coherent state $|\alpha\rangle$ with average photon-number $\langle\hat n\rangle=|\alpha|^2$. As a classical state it fails to qualify for any criterion of the form $g^{(k)}<g^{(k)}_\text{min}<1$. However, we have found for $k=2$ that the effective second-order correlation function $\tilde g^{(2)}$ falls below this boundary for $\langle\hat n\rangle<\ln(2)\approx0.63$. Thus, we concluded that the single-photon criterion is actually independent of the nonclassicality criterion $g^{(2)}<1$ for sub-Poissonian light, as additional information is required to quantify the projection, and this information in turn makes it possible to describe this projection for some classical states. 

In general we note that $g^{(k)}[|\alpha\rangle]=1$ for all $k$ and
\begin{equation}
	\tilde g^{(k)}=(1-e^{-|\alpha|^2})^{k-1}<\frac{k!}{k^k}
\end{equation}
is the condition for a nonzero sub-$k$ projection with our criteria. Using Stirling's approximation for the factorial, we also find a large-$k$ approximation of
\begin{equation}
	|\alpha|^2<1-\ln(e-1)\approx0.46.
\end{equation}
This is again a lower bound for all $k$, meaning that also the general statement of a nonzero sub-$\tilde k$ projection is not a definite nonclassicality criterion, just lies within the range of the nonclassicality criterion $g^{(k)}<1$. All coherent states with average photon number below $0.46$ can be analyzed by our refined criterion.

In comparison, for a thermal state 
\begin{equation}
	\hat\varrho_\text{th}=(1-\lambda)\sum\limits_{j=0}^\infty\lambda^j|j\rangle\langle j|,\quad \lambda\in[0;1[
\end{equation}
with $\langle\hat n\rangle=\lambda/(1-\lambda)$ and $g^{(k)}=k!$, we easily deduce as condition for applying our conditions
\begin{equation}
	\lambda<\frac{1}{k}k^{-\tfrac{1}{k-1}}<\frac{1}{k}.
\end{equation}
While there exists a nonzero lower bound for the excitation of the state, it goes to zero for large k, indicating only very low excited thermal states allow an analysis via our criteria.

\section{Measurement issues}\label{sec.meas}
Setups to determine higher-order correlation functions based on balanced-homodyne correlation measurements were proposed in 2006~\cite{Shchukin2006}. Their experimental validation, performed with the help of waveguide delay lines, established this proposal as a viable method for determining up to $g^{(6)}$~\cite{Silberhorn2010}. Additionally, the vacuum projection of a light field can be directly obtained from click detectors, recording the ratio between no clicks and clicks~\cite{Eisaman2011}. Yet, at least for lower average photon numbers arrays of click detectors already give sufficient information to obtain the photon number statistics and consequently all $g^{(k)}$ and $\tilde g^{(k)}$~\cite{Kroeger2017,Joensson2019}. Combining balanced homodyne correlation measurements with click-detectors is a versatile method to obtain $g^{(k)}$ and $p_0$. Using advanced click-detector arrays may then serve to validate the predictions of this work.

A way to determine $\tilde g^{(2)}$ directly was proposed in~\cite{Hong2017}. Therein, the authors consider a one-to-one optomechanical coupling between an optical photon and a mechanical phonon. Thus single-phonon states could be detected via single-photon measurements, which in turn could be found from Hanbury-Brown Twiss measurement of $g^{(2)}$. To circumvent the problem of strong vacuum components and low signal-to-noise ratio, the authors employed post-selection methods. By first detecting the emission of a photon before actually applying the $g^{(2)}$ measurement they effectively cut out all cases of zero photons. From a theoretical point of view, this generates the effective second-order correlation function $\tilde g^{(2)}$ instead of $g^{(2)}$. The method can be adapted directly for higher-order correlations functions to determine $\tilde g^{(k)}$ without knowledge of the vacuum itself. One major drawback however, is that we lose the information about the vacuum projection of the original quantum state. Hence, the connection to sub-Poissonian light, which was previously drawn, is no longer given. As shown in the applications, even coherent or thermal states may be (correctly) identified as states with sub-$\tilde k$ projection, but not show any nonclassical properties. If such a connection is intended to be established, the original $g^{(k)}$ has to be determined, either by not removing the vacuum, or additionally measuring $p_0$ and computing $g^{(k)}$ from that.

To estimate $p_0$, we may use the knowledge of multiple $g^{(k)}$. Assume that we have no direct information on $p_0$, and that for one $k$ we find $g^{(k+1)}=0$, but $g^{(k)}\neq0$. Hence, the space of Fock states cuts off after $|k\rangle$. Consider again Eq.~(\ref{eq.exactfinal}). Due to the limitation of the super-$k$ space, we know that $N_Q=k$ and $g^{(k)}_Q=g^{(k)}_\text{min}$. This leaves us with the exact equation
\begin{align}
	N_PP=&k\left[\sqrt[k]{\frac{g^{(k)}_\text{min}}{g^{(k)}}Q}-Q\right].
\end{align}
In this case we can use the lower bound on the left-hand side as $N_PP\geq \tilde P$ to find an upper bound on the sub-$\tilde k$ space as
\begin{align}
	\tilde P\leq&k\left[\sqrt[k]{\frac{g^{(k)}_\text{min}}{g^{(k)}}Q}-Q\right].
\end{align}
As $Q$ has the upper bound $Q_\text{max}$, there is a nonzero lower bound for $p_0$ if $\tilde P+Q_\text{max}<1$. This allows us then to find an even smaller $Q_\text{max}$ due to a nonzero vacuum and iteratively approach the correct value for $p_0$. On one hand the connection to the necessity of more information than $g^{(k)}$ to determine the sub-$\tilde k$ projection is obvious. On the other hand, this also ties into the notion of the alternative measure for single-photon sources based on the detection filtering in~\cite{Laussy16}. There, the authors combined the information of different $g^{(k)}$ ($k\geq2$) to define a norm which analyses the sub-Poissonian character to different orders simultaneously.

\section{Conclusions}\label{sec.concl}
We have studied the relation between the $k$th-order correlation function $g^{(k)}$ and the projection of the underlying quantum state of light onto different subspaces. $g^{(k)}$ is a quasiconcave function, from which we conclude that for $0<g^{(k)}<g^{(k)}[|k\rangle]$ there is a nonzero projection on the sub-$k$ space, the sub-$\tilde k$ space and the super-$k$ space. It is possible to give an explicite nonzero lower bound for the first, but not the latter two. The value of $g^{(k)}$ gets artificially enhanced by vacuum. By introducing the effective $k$th-order correlation function $\tilde g^{(k)}$ we account for this vacuum effect. With $\tilde g^{(k)}$, a lower bound for the ratio of the sub-$\tilde k$ projection to super-$k$ projection follows, and an optimized version of the lower bound for the sub-$k$ projection. Including the vacuum projection as an additional information allows to quantify also the sub-$\tilde k$ projection. However, this approach reveals that the connection between $k$th-order correlation function and sub-$\tilde k$ space is independent of nonclassicality. We showed that there is a large-$k$ approximation which is a valid lower bound for all $k$. Finally, we presented some examples of states to apply our criteria for and discussed the measurability of $\tilde g^{(k)}$.

Our results open up a different view and possibly a different field in optical physics. Up to this point, higher-than-second-order correlation functions have been used exclusively for identifying quantum phenomena. In contrast, $g^{(2)}$ has already been established as a source for various information beyond just detecting sub-Poissonian or antibunched light. This work gives insight into a new application of higher-order correlation functions, which at face value appears quantum, but in hindsight is independent of nonclassical phenomena.

\section*{Acknowledgments}
The author acknowledges fruitful discussions with Blas Manuel Rodr\'iguez-Lara. 
This work was supported by the EU through the H2020-FETOPEN grant No. 800942 640378 (ErBeStA), and by the
DNRF through the through the Thomas Pohl Professorship maQma.


%

\end{document}